\documentclass[a4paper,11pt]{article}
\usepackage{pos}

\hypersetup{pdfauthor={Reinhold Kaiser},
            pdftitle={Approaching the Continuum Limit of the Deconfinement Critical Point for \texorpdfstring{$\nf=2$}{Nf=2} Staggered Fermions},
            pdfsubject={QCD Phase Diagram},
            pdfkeywords={LQCD, Deconfinement Phase Transition, Columbia Plot},
            pdfcreator={Latex with hyperref},
}
\usepackage{soul}
\usepackage{mdframed}
\usepackage{tikz}
\usetikzlibrary{
	shapes.geometric,
	arrows, backgrounds,
	positioning,
	calc,
	patterns,
}
\usepackage{pgfplots}
\usepackage{pgfplotstable}
\usepgfplotslibrary{fillbetween}
\pgfplotsset{
compat=1.16,
cmhplot/.style={color=black,mark=none,line width=1pt,<->},
soldot/.style={color=black,only marks,mark=*},
every tick label/.append style={font=\tiny},
legend style={font=\tiny},
/pgf/declare function={
		l(\m,\ns,\a,\mc)=1.604+\a*(1/\m-1/\mc)*\ns^(1/0.6301);
	},
	/pgf/declare function={
		lc(\m,\ns,\a,\mc,\b)=(1.604+\a*(1/\m-1/\mc)*\ns^(1/0.6301))*(1+\b*\ns^(-0.8940));
	},
	select coords between index/.style 2 args={
    	x filter/.code={
        	\ifnum\coordindex<#1\fi
        	\ifnum\coordindex>#2\fi
    	}
	},
}

\usepackage{wrapfig}
\usepackage{physics}
\usepackage{bbold}
\usepackage{csquotes}
\usepackage{bm}
\usepackage{subcaption}

\newcommand{\clqcd}{\texttt{CL\kern-.25em\textsuperscript{2}QCD}}
\newcommand{\nf}{N_\text{f}}
\newcommand{\betapc}{\beta_{\text{pc}}}
\newcommand{\betac}{\beta_{\text{c}}}
\newcommand{\amc}{am_{\text{c}}}

\title{Approaching the Continuum Limit of the Deconfinement Critical Point for $N_\text{f}{=}2$ Staggered Fermions}
\ShortTitle{The Deconfinement Critical Point for $N_\text{f}{=}2$ Staggered Fermions}

\author*[a,b]{Reinhold Kaiser}
\author[a,b]{Owe Philipsen}

\affiliation[a]{Institute for Theoretical Physics - Goethe University,\\
  Max-von-Laue-Str. 1, 60438 Frankfurt am Main, Germany}

\affiliation[b]{John von Neumann Institute for Computing (NIC), GSI,\\
Planckstr. 1, 64291 Darmstadt, Germany}

\emailAdd{kaiser@itp.uni-frankfurt.de}
\emailAdd{philipsen@itp.uni-frankfurt.de}

\abstract{Quenched QCD at zero baryonic chemical potential undergoes a first-order deconfinement phase
transition at a critical temperature $T_c$, which is related to the spontaneous breaking of the global center symmetry.
The center symmetry is broken explicitly by including  dynamical quarks, which weaken the first-order phase transition for decreasing quark masses.
At a certain critical quark mass, which corresponds to the $Z(2)$-critical point, the first-order phase transition turns into a smooth crossover.
We investigate the $Z(2)$-critical quark mass for $\nf=2$ staggered fermions on $N_\tau=8, 10$ lattices, where larger $N_\tau$ correspond to finer lattices.
Monte-Carlo simulations are performed for several quark mass values and aspect ratios in order to extrapolate to the thermodynamic limit.
We present final results for $N_\tau=8$ and preliminary results for $N_\tau=10$ for the critical mass, which are obtained from fitting to a kurtosis finite size scaling formula of the absolute value of the Polyakov loop.
}

\FullConference{%
  FAIR next generation scientists - 7th Edition Workshop (FAIRness2022)\\
  23-27 May 2022\\
  Paralia (Pieria, Greece)
}

\begin{document}
\maketitle

\section{Introduction}
Quantum Chromodynamics (QCD) is the fundamental theory of the strong interaction, whose elementary degrees of freedom are the quarks and gluons.
At low temperatures, the coupling strength between the quarks and gluons is large, confining them to bound hadron states.
At high temperatures, the coupling strength is weak and the constituents are expected to form the so-called quark-gluon plasma.
The various forms of matter are depicted in the QCD phase diagram as a function of temperature and baryonic chemical potential.
The hadronic phase in the QCD phase diagram is separated from the quark-gluon plasma by a transition line, whose nature has to be determined by first-principle calculations and experiments.
This work uses the approach of lattice QCD Monte-Carlo simulations to determine the nature of the deconfinement transition.
However, a severe sign problem prohibits the application of Monte-Carlo importance sampling methods to strongly interacting thermodynamic systems with non-zero real chemical potential $\mu$.
Therefore, studying the theory as a function of the remaining parameters (temperature, quark masses and purely imaginary chemical potential), also for unphysical values, becomes more important.

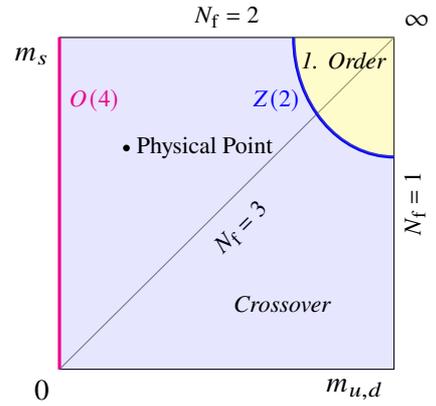
\begin{wrapfigure}[15]{r}{0.39\textwidth}
\begin{tikzpicture}[scale=0.44]
\def\a{10}
\def\b{10}
\def\centerarc[#1](#2)(#3:#4:#5)
    { \draw[#1] ($(#2)+({#5*cos(#3)},{#5*sin(#3)})$) arc (#3:#4:#5 and #5*1.2 ); }
\coordinate (LB) at (0,0);
\coordinate (CB) at (\a/2,0);
\coordinate (RB) at (\a,0);
\coordinate (RC) at (\a,\b/2);
\coordinate (RT) at (\a,\b);
\coordinate (CT) at (\a/2,\b);
\coordinate (LT) at (0,\b);
\coordinate (LC) at (0,\b/2);
\coordinate (chiralB) at (\a/12*5,0);
\coordinate (chiralT) at (\a/32,\b);
\coordinate (PP) at (\a/5,\b/3*2);

\draw (LB) -- (RB) -- (RT) -- (LT) -- cycle;
\draw[gray] (LB) -- node[label={[rotate=45, color=black] below:\footnotesize{$\nf=3$}}]{} (RT);
\centerarc[blue,very thick, name path=C](RT)(180:270:\a*0.3);
\path[name path=BB](RB) -- (LB) -- (LT);
\path[name path=A] (chiralT) -- (LT) -- (LB) -- (chiralB);
\path[name path=D] (3/4*\a, \a) -- (RT) -- (\a, \a/4*3);
\tikzfillbetween[of=C and BB]{blue, opacity=0.1};
\tikzfillbetween[of=C and D]{yellow, opacity=0.2};
\draw[magenta,very thick, name path=B] (LT) -- (LB);
\node[anchor=north west, color=magenta] at (0,\b/8*7) {\footnotesize{$O(4)$}};

\node[anchor=north east] at (LB) {$0$};
\node[anchor=south west] at (RT) {$\infty$};
\node[anchor=north east] at (LT) {$m_s$};
\node[anchor=north east] at (RB) {$m_{u,d}$};
\node[anchor=south] at (CT) {\footnotesize{$\nf=2$}};
\node[anchor=north, rotate=90] at (RC) {\footnotesize{$\nf=1$}};
\fill[black] (PP) circle (\a/100);
\node[anchor=west] at (PP) {\footnotesize Physical Point};
\node[anchor=north] at (\a/3*2, \b/4) {\footnotesize\textit{Crossover}};
\node[anchor=north east, color=blue] at (\a/48*36,\b/24*21) {\footnotesize $Z(2)$};
\node[anchor=south east] at (\a*129/128,\b/16*14) {\footnotesize\textit{1. Order}};

\end{tikzpicture}
\caption{The Columbia plot for the second-order scenario of the chiral phase transition~\cite{cuteri21}. \label{fig:columbia}}
\end{wrapfigure}

Investigating the QCD thermal transition at $\mu=0$ as a function of the three lightest quark masses, the degenerate $u$- and $d$-quarks and the $s$-quark, leads to the Columbia plot of QCD (see fig.~\ref{fig:columbia}).
At physical quark masses the QCD thermal transition is an analytic, smooth crossover~\cite{aoki06}.
In contrast to the light quark mass region, where strong evidence exists, that there is no first-order region in the continuum limit~\cite{cuteri21}, the first-order region in the heavy mass region is known to persist from investigations of pure gauge theory~\cite{boyd96}.

This work focuses on the deconfinement transition in the heavy quark mass regime, which is related to the spontaneous breaking of the of the $Z(3)$ center symmetry.
In the limit of infinite quark masses the $Z(3)$ center symmetry of QCD is exact.
Including heavy, dynamical quarks breaks the center symmetry explicitly, such that decreasing quark masses weaken the first-order phase transition. At the $Z(2)$ second-order boundary it turns into a smooth, analytic crossover.

This $Z(2)$-critical point has been investigated for $\nf=2$ quark flavors and three different lattice spacings employing the Wilson fermion action~\cite{cuteri20}.
The goal is to locate the same $Z(2)$-critical point in the continuum limit employing the unimproved staggered fermion action.
Continuing the work from ref.~\cite{kaiser22}, we present results for the critical quark mass for two different lattice spacings, that are final for the coarser lattice and preliminary for the finer lattice.
The results will provide a first-principles benchmark for effective theories, that are not limited by the sign problem at non-zero real chemical potential.
These effective theories include effective lattice theories obtained from the hopping parameter expansion~\cite{fromm12,saito14,aarts16} and effective Polyakov loop theories in the continuum~\cite{fischer15,lo14}.
Furthermore, investigating QCD for heavy quarks offers the opportunity to study the interplay between dynamical screening, which happens in vaccuum as well as in medium, and Debye screening, which only happens at finite temperature.

\section{Simulation Parameters}
To perform Monte Carlo simulations, a discretized form of the continuum QCD action on a Euclidean $3{+}1$D lattice with dimensions $N_\sigma^3{\times} N_\tau$ is used.
For the gauge sector, the standard Wilson gauge action is used, introducing the inverse gauge coupling $\beta$ as a parameter.
For the fermion sector, the staggered fermion action depends on the quark mass parameter $am$.
The detailed formulations of the staggered fermion and Wilson gauge action can be found in ref.~\cite{gattringer10}.
$\beta(a)$ implicitly tunes the lattice spacing $a$.
Here, it is used to also tune the temperature $T=\frac{1}{a(\beta)N_\tau}$.
In order to approach the continuum limit, $N_\tau$ is increased while $T$ stays constant.
We perform simulations at $N_\tau=8,10$.
For the thermodynamic limit, five different aspect ratios $LT=N_\sigma/ N_\tau\in\{4,5,6,7,8\}$ are simulated.
Localizing the transition point requires to simulate at 2 to 4 $\beta$-values around the pseudo-critical $\betapc$.
The quark mass $am$ is tuned around the critical mass $\amc$ at the $Z(2)$-critical point, simulating 5 to 6 different values.
After a sufficient amount of thermalization steps, 4 independent Markov chains are produced for each set of parameters to increase statistics.

The RHMC algorithm is employed, which is implemented by \clqcd, an open source, Open-CL based lattice QCD code~\cite{sciarra21a}.
Its executable \texttt{rhmc} is run on the GPU-clusters L-CSC at GSI in Darmstadt and on the Goethe-HLR at the Center for Scientific Computing in Frankfurt.
The bash tool \texttt{BaHaMAS} is used to submit and monitor those huge amounts of simulations efficiently~\cite{sciarra21b}.

\section{Analysis of the Deconfinement Transition}
The order parameter associated with the deconfinement transition is the expectation value of the Polyakov loop $L$, averaged over the spatial lattice sites,
\begin{equation}
L =\frac{1}{N_\sigma^3}\sum_{\bm n}\frac{1}{3}\Tr\left[\prod_{n_4=0}^{N_\tau-1}U_4(\bm n, n_4)\right].
\end{equation}
In order to localize the deconfinement transition and determine its order, we analyze the skewness, $B_3$, and the kurtosis, $B_4$, of the norm of the Polyakov loop, which are the third and fourth standardized moments.
The definition of the standardized moments of the norm of the Polyakov loop is given by
\begin{equation}
B_n = \frac{\expval{\left(\abs{L}-\expval{\abs{L}}\right)^n}}{\expval{\left(\abs{L}-\expval{\abs{L}}\right)^2}^{n/2}}.
\end{equation}
The symmetry condition of the distribution of the Polyakov loop $B_3(\betapc)=0$ determines the pseudo-critical $\betapc$.
In the infinite volume limit $B_4(\betapc)$ assumes universal values that are specific to the order of the transition:
$B_4(\betapc)=1$ for a first-order phase transition, $B_4(\betapc)=3$ for a crossover and $B_4(\betac)=1.604(1)$~\cite{blote95} for a $Z(2)$ second-order phase transition in three dimensions.
The coarse sampling of the simulated $\beta$ values requires the interpolation of the data for the skewness and the kurtosis using the multiple histogram reweighting method~\cite{ferrenberg89}.
Good estimates for $\betapc$ and $B_4(\betapc)$ can be extracted from the reweighted data.

Since the measured values for $B_4(\betapc, N_\sigma, am)$ are volume dependent, the finite size scaling formula
\begin{equation}\label{equ:kurtosis-finite-size}
B_4(\betapc,N_\sigma, am)=\left(1.604+B x + 
\order{x^2}\right)\cdot \left( 1 + C N_\sigma^{y_t-y_h} + \order{N_\sigma^{2(y_t{-}y_h)}}\right)
\end{equation}
relates those values to the infinite volume kurtosis value of $1.604$ with constants $B,C$~\cite{takeda16}.
The scaling variable $x=\left(\frac{1}{am}-\frac{1}{\amc}\right)N_\sigma^{1/\nu}$ and the correction term $CN_\sigma^{y_t-y_h}$ depend on the known $Z(2)$-critical exponents $y_t=1/\nu=1.5870(10)$ and $y_h=2.4818(3)$~\cite{pelissetto02}.
For sufficiently large volumes the correction term can be neglected.
By fitting equation \eqref{equ:kurtosis-finite-size} to the $B_4(\betapc,N_\sigma, am)$ data for a certain $N_\tau$, the critical mass $\amc$ can be extracted as a fit parameter.

\section{Numerical Results}
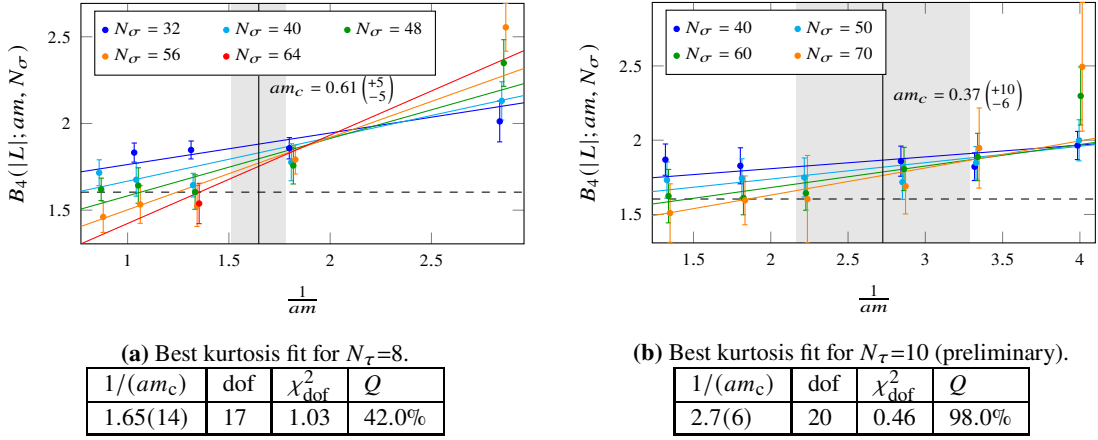
\begin{figure}
\captionsetup[subfigure]{justification=centering}
\centering
\begin{subfigure}[b]{0.49\textwidth}
\centering
\begin{tikzpicture}[every mark/.append style={mark size=1pt}, every error bar/.append style={mark size=1pt}]
\pgfplotstableread{anc/kurtosis-points-nt8.dat}{\kurtosistable};
\begin{axis}[width=\textwidth, height=4.77cm, enlargelimits=false, legend pos=north west, legend columns=3, legend style={/tikz/every even column/.append style={column sep=0.5cm}}, xlabel=\scriptsize $\frac{1}{am}$, ylabel={\scriptsize $B_4(\abs{L}; am, N_\sigma)$}, clip=false]
\path[fill=black,draw=none, opacity=0.1]
(axis cs:1.783,\pgfkeysvalueof{/pgfplots/ymin}) --
(axis cs:1.783,\pgfkeysvalueof{/pgfplots/ymax}) -- 
(axis cs:1.512,\pgfkeysvalueof{/pgfplots/ymax}) --
(axis cs:1.512,\pgfkeysvalueof{/pgfplots/ymin}) -- cycle;
\addplot [black, dashed, domain=1/1.15-0.1:1/0.35+0.1, forget plot]{1.604}; 
\addplot [
	blue,
	select coords between index={0}{3},
	only marks,
	mark size=0.5mm,
	error bars/.cd, y dir = both, y explicit,
] table [x expr=1/\thisrowno{0}-0.02, y=Kurtosis, y error=errorKurtosis] {\kurtosistable};
\addplot [
	cyan,
	select coords between index={4}{8},
	only marks,
	mark size=0.5mm,
	error bars/.cd, y dir = both, y explicit,
] table [x expr=1/\thisrowno{0}-0.01, y=Kurtosis, y error=errorKurtosis] {\kurtosistable};
\addplot [
	green!60!black,
	select coords between index={9}{13},
	only marks,
	mark size=0.5mm,
	error bars/.cd, y dir = both, y explicit,
] table [x expr=1/\thisrowno{0}, y=Kurtosis, y error=errorKurtosis] {\kurtosistable};
\addplot [
	orange,
	select coords between index={14}{18},
	only marks,
	mark size=0.5mm,
	error bars/.cd, y dir = both, y explicit,
] table [x expr=1/\thisrowno{0}+0.01, y=Kurtosis, y error=errorKurtosis] {\kurtosistable};
\addplot [
	red,
	select coords between index={19}{19},
	only marks,
	mark size=0.5mm,
	error bars/.cd, y dir = both, y explicit,
] table [x expr=1/\thisrowno{0}+0.02, y=Kurtosis, y error=errorKurtosis] {\kurtosistable};
\legend{$N_\sigma=32$, $N_\sigma=40$, $N_\sigma=48$, $N_\sigma=56$, $N_\sigma=64$};
\addplot [blue, domain=1/1.15-0.1:1/0.35+0.1]{lc(1/x, 32, 0.0006341, 0.607, 3.827)};
\addplot [cyan, domain=1/1.15-0.1:1/0.35+0.1]{lc(1/x, 40, 0.0006341, 0.607, 3.827)};
\addplot [green!60!black, domain=1/1.15-0.1:1/0.35+0.1]{lc(1/x, 48, 0.0006341, 0.607, 3.827)};
\addplot [orange, domain=1/1.15-0.1:1/0.35+0.1]{lc(1/x, 56, 0.0006341, 0.607, 3.827)};
\addplot [red, domain=1/1.15-0.1:1/0.35+0.1]{lc(1/x, 64, 0.0006341, 0.607, 3.827)};

\draw (axis cs:1.6474283086845711,\pgfkeysvalueof{/pgfplots/ymin}) -- (axis cs:1.6474283086845711,\pgfkeysvalueof{/pgfplots/ymax});
\node[anchor = west] at (axis cs: 1.65, 2.2) {\tiny $am_c=0.61\left(^{+5}_{-5}\right)$};

\end{axis}
\end{tikzpicture}
\caption{Best kurtosis fit for $N_\tau{=}8$.\\
\begin{tabular}{|l|l|l|l|}
\hline
$1/(\amc)$ &  $\text{dof}$ & $\chi^2_{\text{dof}}$ & $Q$\\
\hline
$1.65(14)$ & $17$ & $1.03$ & $42.0\%$\\
\hline
\end{tabular}
}
\end{subfigure}
\hfill
\begin{subfigure}[b]{0.49\textwidth}
\begin{tikzpicture}[every mark/.append style={mark size=1pt}, every error bar/.append style={mark size=1pt}]
\pgfplotstableread{anc/kurtosis-points-nt10.dat}{\kurtosistable};
\begin{axis}[width=\textwidth, height=4.77cm, enlargelimits=false, legend pos=south east, legend columns=2, legend style={/tikz/every even column/.append style={column sep=0.5cm}}, xlabel=\scriptsize $\frac{1}{am}$, ylabel={\scriptsize $B_4(\abs{L}; am, N_\sigma)$}, clip=false, legend pos=north west]
\path[fill=black,draw=none, opacity=0.1]
(axis cs:3.286688871,\pgfkeysvalueof{/pgfplots/ymin}) --
(axis cs:3.286688871,\pgfkeysvalueof{/pgfplots/ymax}) -- 
(axis cs:2.162710442,\pgfkeysvalueof{/pgfplots/ymax}) --
(axis cs:2.162710442,\pgfkeysvalueof{/pgfplots/ymin}) -- cycle;
\addplot [black, dashed, domain=1/0.75-0.1:1/0.25+0.1, forget plot]{1.604}; 
\addplot [
	blue,
	select coords between index={0}{4},
	only marks,
	mark size=0.5mm,
	error bars/.cd, y dir = both, y explicit,
] table [x expr=1/\thisrowno{0}-0.015, y=Kurtosis, y error=errorKurtosis] {\kurtosistable};
\addplot [
	cyan,
	select coords between index={5}{10},
	only marks,
	mark size=0.5mm,
	error bars/.cd, y dir = both, y explicit,
] table [x expr=1/\thisrowno{0}-0.005, y=Kurtosis, y error=errorKurtosis] {\kurtosistable};
\addplot [
	green!60!black,
	select coords between index={11}{16},
	only marks,
	mark size=0.5mm,
	error bars/.cd, y dir = both, y explicit,
] table [x expr=1/\thisrowno{0}+0.005, y=Kurtosis, y error=errorKurtosis] {\kurtosistable};
\addplot [
	orange,
	select coords between index={17}{22},
	only marks,
	mark size=0.5mm,
	error bars/.cd, y dir = both, y explicit,
] table [x expr=1/\thisrowno{0}+0.015, y=Kurtosis, y error=errorKurtosis] {\kurtosistable};
\legend{$N_\sigma=40$, $N_\sigma=50$, $N_\sigma=60$, $N_\sigma=70$};
\addplot [blue, domain=1/0.75-0.1:1/0.25+0.1]{lc(1/x, 40, 0.0001958559, 1/2.724699656, 4.414235294)};
\addplot [cyan, domain=1/0.75-0.1:1/0.25+0.1]{lc(1/x, 50, 0.0001958559, 1/2.724699656, 4.414235294)};
\addplot [green!60!black, domain=1/0.75-0.1:1/0.25+0.1]{lc(1/x, 60, 0.0001958559, 1/2.724699656, 4.414235294)};
\addplot [orange, domain=1/0.75-0.1:1/0.25+0.1]{lc(1/x, 70, 0.0001958559, 1/2.724699656, 4.414235294)};

\draw (axis cs:2.724699656,\pgfkeysvalueof{/pgfplots/ymin}) -- (axis cs:2.724699656,\pgfkeysvalueof{/pgfplots/ymax});
\node[anchor = west] at (axis cs: 2.724699656, 2.3) {\tiny $am_c=0.37\left(^{+10}_{-6}\right)$};

\end{axis}
\end{tikzpicture}
\caption{Best kurtosis fit for $N_\tau{=}10$ (preliminary).\\
\begin{tabular}{|l|l|l|l|}
\hline
$1/(\amc)$ & $\text{dof}$ & $\chi^2_{\text{dof}}$ & $Q$\\
\hline
$2.7(6)$  & $20$ & $0.46$ & $98.0\%$\\
\hline
\end{tabular}
}
\centering
\end{subfigure}
\caption{Kurtosis data and fits for $N_\tau=8,10$.
The data points are shifted due to readability. The colored lines show the combined fit, indicating the corresponding volume by the color.
The dashed line indicates the infinite volume kurtosis value for the $Z(2)$ second-order transition.
The vertical black line and the grey band localize the critical mass and its error. 
 \label{fig:kurtosis-fits}}
\end{figure}

\begin{wrapfigure}[16]{r}{0.48\textwidth}
\begin{tikzpicture}
    \begin{axis}[
      name=ax1,
      xticklabel style={/pgf/number format/fixed},
      xmin=0, xmax=0.13,
      ymin=3.85, ymax=5.1,
      x dir=reverse,
      width=4.2cm,
      height=5cm,
      axis x line=box,
      axis y line=left,
      x axis line style={-},
      extra x ticks={0.06915, 0.0880},
      extra x tick labels={10, 8},
      extra x tick style={tick pos=top, ticklabel pos=top},
      xtick pos=left,
      legend style={at={(axis cs:0.058, 4.7)}, anchor=south},
      ylabel=\small $m_\pi^{Z(2)}\; \{GeV\}$,
    ]
      \addplot[
  red, mark options={red}, only marks, mark size=0.3ex,
  error bars/.cd, 
    y fixed,
    y dir=both, 
    y explicit
] table [x=x, y=y,y error=error, col sep=comma, row sep=crcr] {
    x,  y,       error\\
    0.0691, 3.95, 0.04\\
    0.0885, 4.01, 0.04\\
    };
      \addlegendimage{color=red,mark=*,only marks, mark size=0.2ex};
      \addlegendentry{\small staggered};
      \addplot [black] coordinates{(0,0) (0,6)};
    \end{axis}
    
    \begin{axis}[
      xticklabel style={/pgf/number format/fixed},
      width=4.2cm,
      height=5cm,
      xmin=0, xmax=0.13,
      ymin=3.85, ymax=5.1,
      yticklabel pos=right,
      at={(ax1.south east)},
      anchor=south west,
      axis x line=box,
      axis y line= right,
      x axis line style={-},
      extra x ticks={0.0691, 0.0876, 0.1186},
      extra x tick labels={10, 8, 6},
      extra x tick style={tick pos=top, ticklabel pos=top},
      xtick pos=left,
      legend style={at={(axis cs:0.06, 3.87)}, anchor=south},
    ]
      \addplot[
  blue, mark options={blue}, only marks, mark size=0.3ex,
  error bars/.cd, 
    y fixed,
    y dir=both, 
    y explicit
] table [x=x, y=y,y error=error, col sep=comma, row sep=crcr] {
    x,  y,       error\\
    0.0691, 4.39, 0.05\\
    0.0876, 4.51, 0.05\\
    0.1186, 5.01, 0.05\\
    };
    \addlegendentry{\small Wilson};
    \end{axis}
    \node[anchor=north, yshift=-1.5ex] at (ax1.south east) {\small $a\; \{\text{fm}\}$};
    \node[anchor=south, yshift=1.5ex] at (ax1.north east) {\small $N_\tau$};
\end{tikzpicture}
\caption{Comparison with the results from Wilson fermions~\cite{cuteri20}. The staggered value for $m_\pi^{Z(2)}$ is preliminary for $am{=}0.35$.}\label{fig:wilson-comparison}

\end{wrapfigure}
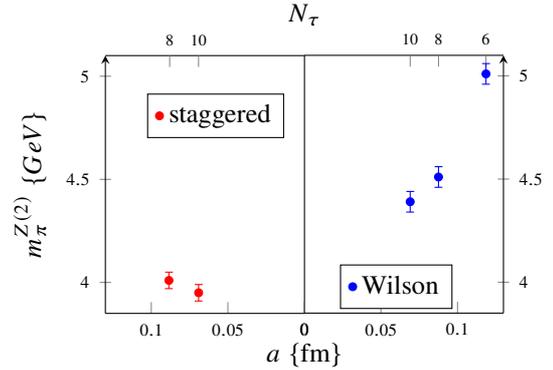

The best kurtosis fits for $N_\tau=8,10$ are shown in fig.~\ref{fig:kurtosis-fits}, where in both cases the correction term from equation~\eqref{equ:kurtosis-finite-size} is relevant.
The effect is pairwise different crossing points of the kurtosis lines for different volumes, that are shifted upwards with respect to the infinite volume kurtosis crossing point (crossing of the black solid and black dashed line).
It can be observed, that the error bars on the kurtosis points for $N_\tau=10$ are larger than for $N_\tau=8$.
Explanations are significantly larger autocorrelation times for the larger lattices and low statistics for the larger $N_\sigma$ for $N_\tau=10$.
The large $Q$-parameter for the fit quality and the large error on the critical mass are a consequence of the large error bars.
With increasing statistics this issue will disappear.

To compare the results, the scale is set via the $w_0$ scale~\cite{borsanyi12} based on the Wilson flow~\cite{luescher10}.
The measurement of the pion mass in physical units allows a comparison with results from Wilson fermions~\cite{cuteri20}, which can be seen in fig.~\ref{fig:wilson-comparison}.
The error bars represent the error from the pion mass measurement. 
The fit error on $\amc$, which is not included, is significantly larger.
The pion mass in lattice units for $N_\tau=8$ is $am_\pi^{Z(2)}=1.79918(9)$ and for $N_\tau=10$ it is $am_\pi^{Z(2)}=1.38124(11)$, implying that the pion, represented by its Compton wavelength, is not resolved by the lattice.

In future, the statistics for $N_\tau=10$ will increase with the currently running simulations, leading to a more reliable result for the critical mass.
For a continuum extrapolation, simulations on finer lattices are required, making it more difficult to obtain results with the same precision due to larger lattices and autocorrelation times.
The discretization effects implied by the large pion masses in lattice units also show the necessity to simulate on finer lattices.


\acknowledgments{
The authors acknowledge support by the Deutsche Forschungsgemeinschaft (DFG, German Research Foundation) through the CRC-TR 211 \enquote{Strong-interaction matter under extreme conditions} – project number 315477589 – TRR 211 and by the State of Hesse within the Research Cluster ELEMENTS (Project ID 500/10.006).
We thank the Helmholtz Graduate School for Hadron and Ion Research (HGS-HIRe) for its support as well as the staff of L-CSC at GSI Helmholtzzentrum für Schwerionenforschung and the staff of Goethe-HLR at the Center for Scientific Computing Frankfurt for computer time and support.
}

%
\bibliographystyle{bibliography/JHEP}
\bibliography{bibliography/bibliography}

\end{document}